\documentclass[10pt,twocolumn]{article}
\usepackage[utf8]{inputenc}
\usepackage[T2A]{fontenc}
\usepackage{amsmath,amssymb}
\usepackage{graphicx}

\title{Geometric Phase of the Two-Particle Bethe Wavefunction}
\author{$^{1}$V.A. Tomilin\footnote{E-mail:~\textsf{8342tomilin@mail.ru}}, $^{1}$A.M. Rostom, $^{1,2}$L.V. Il'ichov
\\
\footnotesize\it $^{1}$Institute of Automation and Electrometry SB RAS, 630090, Novosibirsk, Russia
\\
\footnotesize\it $^{2}$Institute of Laser Physics SB RAS, 630090, Novosibirsk, Russia\\}
\date{}

\begin{document}

\maketitle

\begin{abstract}
We consider a problem of geometric phase generation in a system of two interacting bosons confined in a narrow ring potential with a localized defect. Geometric phase emerges from variation of parameters of the defect. Particle interaction is taken into account within a framework of the Lieb-Liniger model. The energy spectrum is evaluated and its dependence on the parameters of the problem is described. It is shown that the interaction leads to increase of the geometric phase for a given contour of variations. The work is motivated by earlier proposed ideas of quantum gyroscope and quantum accelerometer based on atomic Bose-Einstein condensates.
\end{abstract}

\section{Introduction}

Exactly solvable models in quantum mechanics are sometimes regarded as ``touchstones'' that grant deeper understanding of effects occuring in more complex real systems. In view of rapid development of experimental techniques in the physics of ultra-cold atoms, exactly solvable many-body problems become highly relevant. Optical and magneto-optical traps now allow creation and effective control of low-dimensional spatial configurations of atoms \cite{Bloch,Cazalilla}. Regarding this, any theoretical results obtained even for relatively simple ``toy models'' may be useful in analyzing experimental data.

Theoretical and experimental investigations of low-dimensional atomic systems yield new fundamental knowledge on physics of many-body quantum correlations \cite{Slavnov,Kinoshita,Imambekov,Olshanii,Kozlowski}, many-body systems behaviour in thermodynamical limit \cite{Dunjko,Lang,Minguzzi} and physics of phase transitions \cite{Esslinger,Gemelke}. In addition, such systems have various potential applications, including inertial sensors analogous to SQUID \cite{Ragole}.

In a series of works \cite{Tomilin1,Tomilin2,Tomilin3} an original concept of quantum gyroscope and quantum accelerometer were developed. The main element of those schemes sensitive to non-inertial movement of the reference frame was a pair of ring-shaped modes of atomic Bose-Einstein condensate (BEC), interrupted by the local non-uniformity -- the defect. The structure of this defect is particularly important -- it should define a certain orientation of the ring. The discussed gyroscope and accelerometer thus differ merely by the mutual orientation of the ring-shaped modes. The measured quantity containing information on angular velocity (in case of gyroscope) or linear acceleration (in case of accelerometer) is the difference of geometric phases acquired by the modes after a certain evolution of the parameters of the defect. The downside of the model was the absence of atom-atom interaction, which plays crucial role in low-dimensional systems. In this work we propose to take this interaction into account within the framework of Lieb-Liniger model \cite{LiebLiniger}. It is one of the most popular exactly solvable many-body models, describing a system of bosons with contact repulsive interaction. It is solved by utilizing the Bethe ansatz \cite{Bethe}. The solutions for periodical boundary conditions (which is physically equivalent to ring-like geometry) \cite{LiebLiniger} and for particles in an impenetrable ``box'' \cite{Gaudin} are known. In line with the mentioned proposals of perspective quantum inertial sensors, it is of interest to develop a theory for interacting bosons in one-dimensional ring geometry with a reasonably general assumptions for the defect. The Bethe ansatz was earlier adapted for one-dimensional configurations with simple $\delta$-like defects \cite{Li}. The current work is the first attempt to study the influence of interaction on the generation of geometric phase that emerges when the parameters defining the structure of the defect are varied. To this end, we have considered the simplest two-particle model\footnote{While the Lieb-Liniger model belongs to the class of exactly solvable models, its practical implementation for large number of particles, and especially its generalizations for non-trivial boundary conditions is a difficult task. Nevertheless, even few-body models demonstrate non-trivial effects and may have practical applications, work \cite{Liu} serving as a good example. It studies the dynamics of a pair of atoms initially prepared in a $NOON$-state in a double-well potential divided by a $\delta$-like barrier.}, which still demonstrates the role of interaction in the process of geometric phase generation.

\section{The Model}
\begin{figure}[h!]
  \centering
  \includegraphics[width=0.3\textwidth]{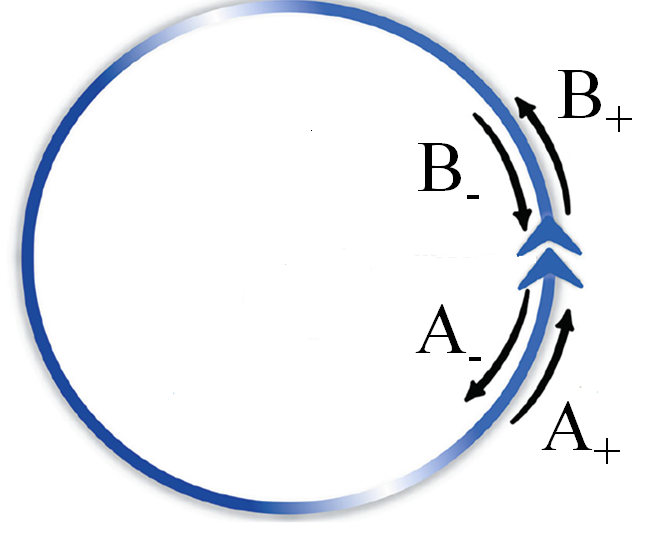}
  \caption{The schematics of the ring-shaped potential with a defect. Transmitted and reflected waves are indicated.}\label{f1}
\end{figure}

We consider a mode of atomic BEC consisting of two particles with momenta $\hat{k}_{1}$ and $\hat{k}_{2}$. The spatial configuration of the mode is the toroid with very small inner radius \footnote{Realization of BEC in such geometry was discussed in \cite{Gupta,Gorlitz}.}. A good initial approximation for such a structure is a one-dimensional ring of length $L$. Positions of the particles are then given by an angular coordinate $x\in[0,L)$. The interaction between the particles is repulsive and point-like, with strength given by the parameter $c$. The Schrodinger equation in $\hbar=m=1$ system of units is then given by 

\begin{equation}
\bigg(\frac{\hat{k}_{1}^{2}}{2}+\frac{\hat{k}_{2}^{2}}{2}\bigg)\Psi(x_{1},x_{2}) +2c\delta(x_{1}-x_{2})\Psi(x_{1},x_{2})=E\Psi(x_{1},x_{2}).
\label{1}
\end{equation}
The problem is then reduced to the two-particle version of the Lieb-Liniger model. However, unlike the well-studied case of  uniform potential, the presence of localized potential defect enforces non-trivial boundary conditions. The plane waves are fundamental solutions of (\ref{1}); in this case, the description of the defect by the transfer matrix formalism is most natural \cite{SanchezSoto}. For simplicity, assume that the scattering properties of the defect are momentum-independent. Transfer matrix $\mathcal{M}$ belongs to the $SU(1,1)$ group and relates single-particle amplitudes of transmitted and reflected plane waves on the two sides of the defect (see Fig. \ref{f1}):

\begin{align}
\begin{split}
&\left(
  \begin{array}{c}
    A_{+} \\
    A_{-} \\
  \end{array}
\right)\doteq\mathcal{M}\left(
  \begin{array}{c}
    B_{+} \\
    B_{-} \\
  \end{array}
\right)=\left(
          \begin{array}{cc}
            u & v \\
            v^{*} & u^{*} \\
          \end{array}
        \right)\left(
  \begin{array}{c}
    B_{+} \\
    B_{-} \\
  \end{array}
\right);
\\
&|u|^{2}-|v|^{2}=1.
\end{split}
\label{2}
\end{align}

Bethe ansatz is based on elementary Bethe wavefunctions \cite{GaudinBook}. They are specified by the momenta of the particles. Since the defect produces reflected waves, both positive and negative momenta are possible. As a result, the elementary  wavefunction in our two-particle model is a superposition of four elementary Bethe wavefunctions:

\begin{align}
\begin{split}
&\Psi(x_{1},x_{2})=\Psi_{\{k_{1},k_{2}\}}(x_{1},x_{2})+ \Psi_{\{k_{1},-k_{2}\}}(x_{1},x_{2})+ 
\\
&\Psi_{\{-k_{1},k_{2}\}}(x_{1},x_{2})+ \Psi_{\{-k_{1},-k_{2}\}}(x_{1},x_{2}),
\end{split}
\label{3}
\end{align}
where the lower indices $\{k,k'\}$ correspond to symmetrized two-particle functions built from plane waves with momenta $k$ and $k'$:

\begin{equation}
\Psi_{\{k_{1},k_{2}\}}(x_{1},x_{2})=\mathop{\sum}_{p=e,\sigma}A^{(p)}(k_{1},k_{2})\cdot \exp[\mathop{\sum}_{j=1,2}\imath k_{j}x_{p(j)}].
\label{4}
\end{equation}
Here $e$ stands for identical permutation, and $\sigma$ stands for $(12)$ permutation.

Each transfer matrix can be represented as a product of ``elementary'' transfer matrices, each of them belonging to one of three possible canonical forms  \cite{SanchezSoto}. We will be using two of them (the same choice was made in  \cite{Tomilin1} and \cite{Tomilin3}):

\begin{equation}
\mathcal{M}=\left(
              \begin{array}{cc}
                e^{\imath\alpha} & 0 \\
                0 & e^{-\imath\alpha} \\
              \end{array}
            \right)\cdot\left(
              \begin{array}{cc}
                \cosh\eta & \imath\sinh\eta \\
                -\imath\sinh\eta & \cosh\eta \\
              \end{array}
            \right).
\label{5}
\end{equation}
The first matrix corresponds to $\alpha$ phase shift of the transmitted waves, while the second one describes the amplitude change, with transmission coefficient equal to $1/\cosh\eta$, and additional $\pi/2$ phase shift for reflected wave. The order of multiplication of ``elementary'' transfer matrices defines the orientation of the ring, which in turn gives a reference frame for assigning positive and negative momenta entering (\ref{3}).

Since the transfer matrix describes a single-particle scattering, obtaining boundary conditions for two-body amplitudes $A^{(p)}(k_{1},k_{2})$ appearing in (\ref{3},\ref{4}) requires additional considerations. The arguments $k_{1},k_{2}$ refer to different particles; it is then clear that the transfer matrix should relate the amplitudes with one of the arguments fixed. For example, let the plane wave amplitudes in the small left neighbourhood of the defect (with respect to the ring orientation) be $A^{(e)}(k_{1},k_{2})$ и $A^{(e)}(-k_{1},k_{2})$. The corresponding amplitude in the right neighbourhood will then differ by phase factors $e^{\pm\imath k_{1}L}$ appearing from circling the ring in a given direction, and also by permutation of the particle coordinates. The scattering of the other particle is treated by simultaneously replacing $\sigma\rightarrow e$ and $k_{1}\rightarrow k_{2}$. This yields the following set of boundary conditions:

\begin{align}
\begin{split}
&\left(
  \begin{array}{c}
    A^{(e)}(k_{1},k_{2}) \\
    A^{(e)}(-k_{1},k_{2}) \\
  \end{array}
\right)=\mathcal{M}\left(
  \begin{array}{c}
    A^{(\sigma)}(k_{1},k_{2})e^{\imath k_{1}L} \\
    A^{(\sigma)}(-k_{1},k_{2})e^{-\imath k_{1}L} \\
  \end{array}
\right),
\\
&\left(
  \begin{array}{c}
    A^{(\sigma)}(k_{1},k_{2}) \\
    A^{(\sigma)}(k_{1},-k_{2}) \\
  \end{array}
\right)=\mathcal{M}\left(
  \begin{array}{c}
    A^{(e)}(k_{1},k_{2})e^{\imath k_{2}L} \\
    A^{(e)}(k_{1},-k_{2})e^{-\imath k_{2}L} \\
  \end{array}
\right).
\end{split}
\label{6}
\end{align}

The boundary conditions for transitions between domains $x_{1}<x_{2}$ and $x_{1}>x_{2}$ in presence of contact interaction have the standard form \cite{GaudinBook}

\begin{equation}
A^{(p)}(k_{1},k_{2})=\bigg(1+\frac{\imath c}{k_{p(1)}-k_{p(2)}}\bigg)A(k_{1},k_{2})
\label{7}.
\end{equation}
Conditions (\ref{6},\ref{7}) constitute a complete system of equations defining two-particle amplitudes. It is consistent if the following spectral relation holds:

\begin{align}
\begin{split}
&(k_{1}^{2}-k_{2}^{2}-c^{2})Re(u\cdot e^{-\imath k_{1}L})-2ck_{1}Im(u\cdot e^{-\imath k_{1}L})=
\\
&k_{1}^{2}-k_{2}^{2}+c^{2},
\\
&(k_{2}^{2}-k_{1}^{2}-c^{2})Re(u\cdot e^{-\imath k_{2}L})-2ck_{2}Im(u\cdot e^{-\imath k_{2}L})=
\\
&k_{2}^{2}-k_{1}^{2}+c^{2}.
\end{split}
\label{8}
\end{align}

\begin{figure}[h!]
  \centering
  \includegraphics[width=0.35\textwidth]{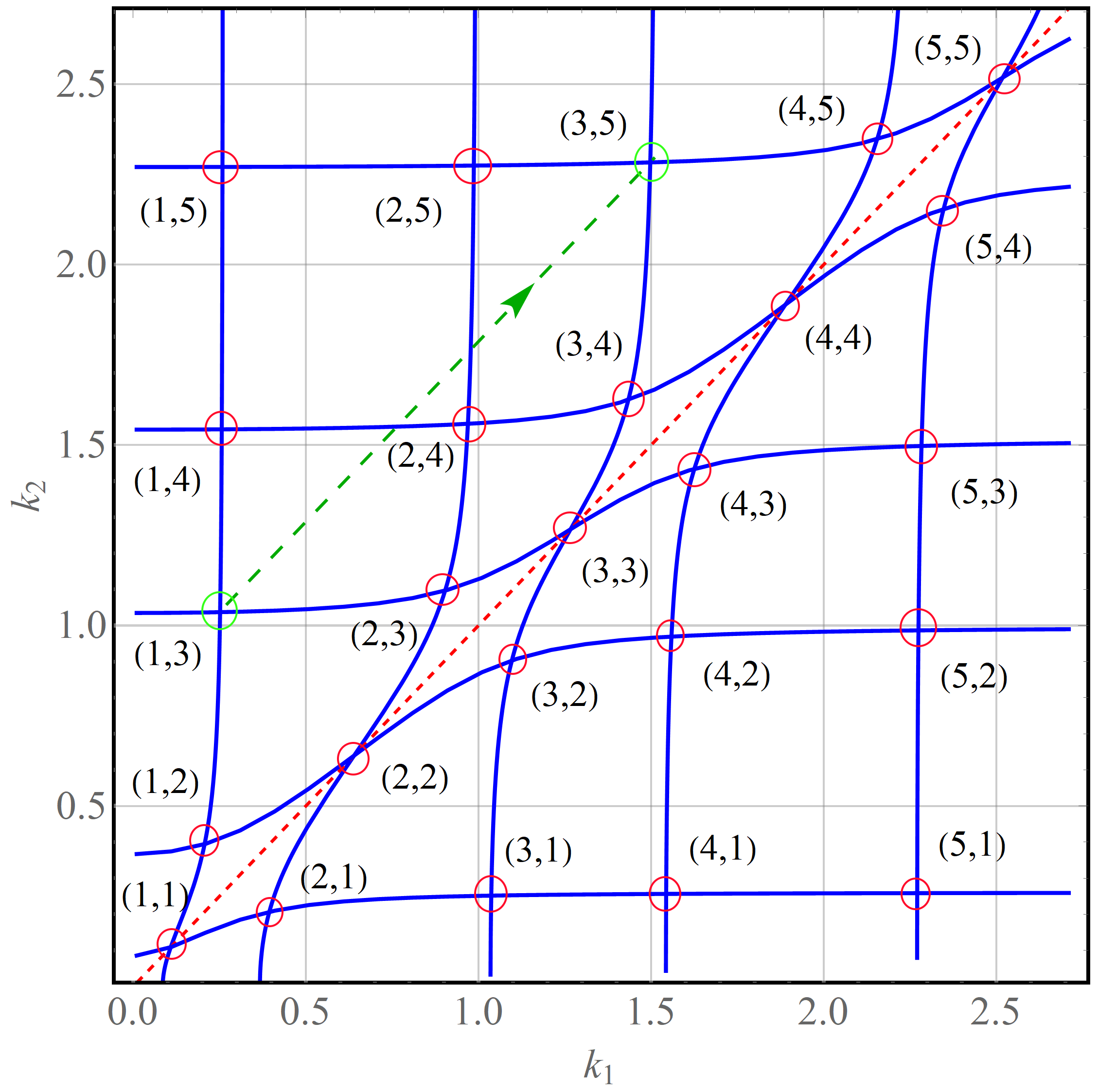}
  \caption{Two sets of curves yielding the solutions of spectral relations for $\eta=2,\alpha=0,c=0.1,L=5$. The roots of the spectral solutions are marked by circles. Each root appears as an intersection vertical and horizontal curve numerated by integers. The dashed line connects two roots with numbers $(1,3)$ and $(3,5)$ transforming into each other upon variation of $\alpha$ by $2\pi$ (see the text).}\label{f2}
\end{figure}

Equations (\ref{8}) describe two sets of curves in $k_{1}k_{2}$ plane -- one ``horizontal'' and one ``vertical'', which are mutually transposed (Fig. \ref{f2}). Each curve can be numerated by an integer number characterizing the proximity to the zero point. The eigenenergies of the problem can thus be enumerated by two indices $(i,j)$ which stand for the numbers of horizontal and vertical curves, the intersection of which form a particular solution of (\ref{1}):

\begin{equation}
E^{(ij)}=\frac{k_{1}^{(i,j)2}}{2}+\frac{k_{2}^{(i,j)2}}{2}.
\label{9}
\end{equation}

In the case of slow variation of parameters $\eta$ and $\alpha$ through some closed contour, the system state gains a phase shift, called geometric phase \cite{Berry}. As a contour of variation, consider the following: $\eta=const$, $\alpha\in[0,2\pi]$. The geometric phase can then be expressed in the following gauge-invariant form \cite{Bohm}:

\begin{align}
\begin{split}
&\theta_{g}=arg\langle\Psi(\alpha=0)|\Psi(\alpha=2\pi)\rangle+
\\
&\frac{\imath}{2}\int_{0}^{2\pi}\frac{\langle\Psi|\frac{\partial}{\partial\alpha}\Psi\rangle- \langle\frac{\partial}{\partial\alpha}\Psi|\Psi\rangle}{\langle\Psi|\Psi\rangle}d\alpha,
\end{split}
\label{10}
\end{align}
with

\begin{equation}
\langle\Phi|\Psi\rangle=\int\Phi^{*}(x_{1},x_{2})\Psi(x_{1},x_{2})dx_{1}dx_{2}.
\label{11}
\end{equation}

Upon variation of $\alpha$, each set of curves defined in (\ref{8}) also changes, but since they only differ by permuting $k_{1}$ and $k_{2}$, this change is the same for ``horizontal'' and ``vertical'' curves. Changing $\alpha$ by the integer of $2\pi$ would evidently bring the system, initially prepared in some steady state, into the other steady state\footnote{While parameters of the defect are varied cyclically, the energy of the is not conserved (by performing work on it). Such a variation can then be used for switching between various steady states.}. Consequently, upon variation of $\alpha$, the solutions of the spectral relations would shift along straight lines $k_{2}=k_{1}+const$ connecting different solutions for $\alpha=0$, as illustrated by Fig. \ref{f2}.

\section{Results}

\begin{figure}[h!]
  \centering
  \includegraphics[width=0.45\textwidth]{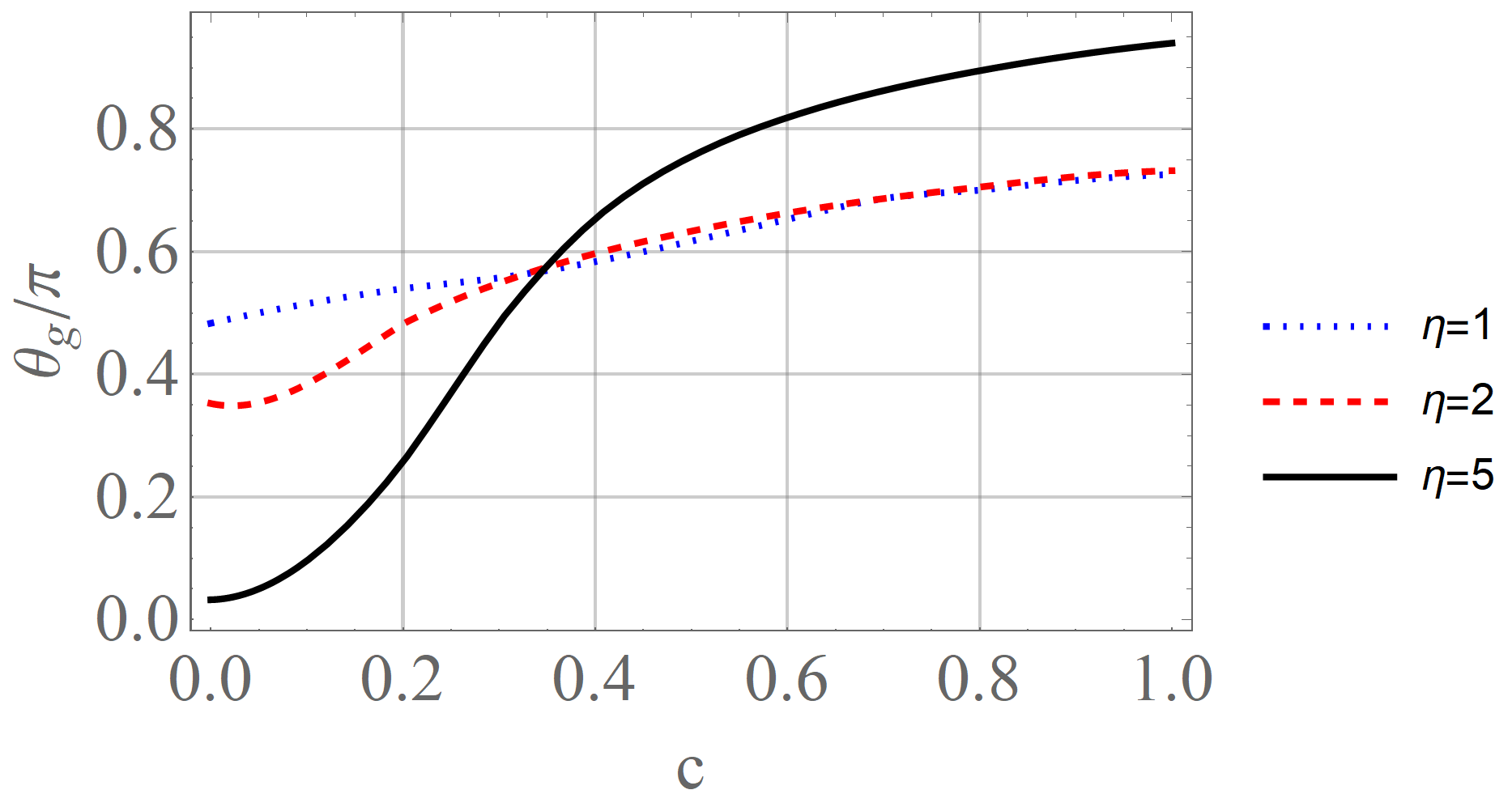}
  \caption{Geometric phase for $(1,3)$ steady state as a function of interaction parameter $c$ for different value of $\eta$. For higher values of $\eta$ and $c$ the geometric phase approaches its maximal asymptotical value of $\pi$. For small values of $\eta$ and $c$ its asymptotic is $\pi/2$.}\label{f3}
\end{figure}

Evaluation of the geometric phase was done for a particular initial state $(1,3)$ (see the specified point in Fig. \ref{f2}). In case of choosing the $(n,n)$ state (and particularly the ground state $(1,1)$), the geometric phase appears to be zero. Fig. \ref{f3} represents the geometric phases as functions of the interaction parameter $c$ for different values of $\eta$. It can be seen that the geometric phase increases monotonically with interaction, approaching some asymptotical value. The system with more reflective defect (corresponding to higher values of $\eta$) produces geometric phase which is more sensitive to the interaction.

These characteristic features appear from the interplay of two effects  -- the influence of the potential defect and the interaction. It is clear that for strong interaction the role of the defect decreases, which follows directly from the spectral relations (\ref{8}) (at least for moderate values of momenta, i.e. for low-lying energy levels). As a result, the dependence of the geometric phase on $c$ is an asymptotical one. The value of this asymptotic increases with $\eta$. The presence of weakly transmitting defect inevitably leads to more pronounced dependence on its parameters in the wavefunctions of the system. Varying these parameters thus leads to increased values of the integral defining the geometric phase.

In the weak interaction regime the situation is reversed  -- more transparent defect produces bigger geometric phase. In the limit of zero interaction the spectral relations (\ref{8}) can be solved analytically:

\begin{equation}
k_{1,2}L=\alpha-\frac{\pi}{2}+\arccos\bigg(\frac{1}{\cosh\eta}\bigg)+2\pi n_{1,2};\,n_{1,2}\in\mathbb{Z}.
\label{12}
\end{equation}
The steady state is then defined by the integer numbers $n_{1,2}$. Two-particle amplitudes with upper indices $e$ and $\sigma$ become equal. As a result, the main contribution to the geometric phase comes from the first term in(\ref{10}), i.e. from the global phase. Its limit for high values of $\eta$ is given by

\begin{equation}
arg\langle\Psi(\alpha=0)|\Psi(\alpha=2\pi)\rangle\rightarrow-\arccos\bigg(\frac{1}{\cosh\eta}\bigg)+\pi/2.
\label{13}
\end{equation}
The geometric phase is then maximized in case of more transparent defect. This result sounds counter-intuitive, but can be explained as follows. Despite the reflection coefficient in the considered case being small, the defect still introduces the phase shifts, as can be seen from (\ref{5}). The dependence of wavefunctions on this phase becomes more pronounced, which leads to the increase of the geometrical phase.

\section{Conclusion}

The effect of atom-atom interaction on the generation of the geometric phase in ring-shaped atomic BEC with a localized potential defect was studied within a framework of a simple two-body model. It was shown that repulsive interaction leads to increase in geometric phase gained by variation of the parameters of the defect. Depending on the interaction strength, it may be preferable to use a defect with either high or low transmission coefficient in order to maximize the geometric phase.

To our best knowledge, the geometric phase for Bethe wavefunctions is investigated for the first time. The scenario of the geometric phase generation was motivated by previously proposed schemes of BEC-based inertial sensors. This makes the obtained results potentially interesting for applications in quantum metrology. For further investigation of its potential prospects, equation (\ref{1}) should be generalized to non-inertial reference frame and solved separately for two modes with differently oriented defects. If the geometric phase difference acquired by the modes and measured through their interference demonstrates the same (albeit less pronounced) dependence on the interaction strength as the one presented in Fig. \ref{f3}, then a positive role of atom-atom interaction could be confirmed. To date, in most standard metrological schemes based on atom interferometers \cite{Dimopoulos,Jannin} the atom-atom interaction is seen as a detrimental factor, worsening overall sensitivity of the scheme.

The complication introduced by the potential defect has limited the current study to the two-particle problem. In case of $N$-body problem, the number of terms in the corresponding equation(\ref{3}) is equal to $2^{N}$. Due to that, the prospects for obtaining complete solution of many-body problem analogous to the classical Bethe ansatz remain unclear.

\textbf{Acknowledgements.} This work was financially supported by the project 23-12-00182 of Russian Science Foundation (https://rscf.ru/project/12-00182/).

\end{document}